\documentclass{svproc}
\usepackage{url}

\usepackage{graphicx}
\begin{document}
\mainmatter              
\title{A statistical analysis of the nuclear structure uncertainties in $\mu$D}
\titlerunning{A statistical analysis of the nuclear structure uncertainties in $\mu$D}  
%
\author{Oscar J. Hernandez\inst{1,2,3} \and Sonia Bacca\inst{1,3,4}  \and Nir Barnea\inst{5} \and Nir Nevo-Dinur\inst{3} \and Andreas Ekstr\"om\inst{6} \and Chen Ji\inst{7} }
\authorrunning{Oscar J. Hernandez et al.} 
%
\tocauthor{Oscar J. Hernandez, Andreas Ekstr\"om, Jeffrey Dean, David Grove,
Craig Chambers, Kim B. Bruce, and Elisa Bertino}
\institute{Institut f\"ur Kernphysik and PRISMA Cluster of Excellence, Johannes-Gutenberg-Universit\"at Mainz,
55128 Mainz, Germany
\email{javierh@phas.ubc.ca},
\and
Department of Physics and Astronomy, University of British Columbia,
Vancouver, BC, V6T 1Z4, Canada,
\and
TRIUMF, 4004 Wesbrook Mall,
Vancouver, BC V6T 2A3, Canada,
\and 
Department of Physics and Astronomy, University of Manitoba,
Winnipeg, MB, R3T 2N2, Canada,
\and 
Racah Institute of Physics, The Hebrew University,
Jerusalem 9190401, Israel
\and 
Department of Physics, Chalmers University of Technology,
SE-412 96 Gothenburg, Sweden
\and 
Institute of Particle Physics, Central China Normal University,
Wuhan 430079, China,
}

\maketitle              

\begin{abstract}

The charge radius of the deuteron (D), was recently determined to three times the precision compared with previous measurements using the measured Lamb shift in muonic deuterium ($\mu$D). However, the $\mu$D value is 5.6 $\sigma$ smaller than the world averaged CODATA-2014 value \cite{Pohl_2016}. To shed light on this discrepancy we analyze the uncertainties of the nuclear structure calculations of the Lamb shift in $\mu$D and conclude that nuclear theory uncertainty is not likely to be the source of the discrepancy.

\end{abstract}
\section{Introduction}
The two-photon exchange (TPE) contribution is a crucial ingredient in the precision determination of the charge radius from Lamb shift (LS) measurements in muonic atoms. The charge radius is extracted from the measurements of the $2S$-$2P$ energy splitting $\rm \Delta E_{\rm LS}$ through
\begin{equation}
{\rm \Delta} E_{\rm LS} = \delta_{\rm QED}+ \delta_{\rm TPE} + \delta_{\rm FS}(r^2_d),\label{eq:two-photon exchange}
\end{equation}
valid up to fifth order in $(Z\alpha)$, where $Z$ is the charge number of the nucleus and $\alpha$ is the fine structure constant. The term $\delta_{\rm QED}$ denote the quantum electrodynamic (QED) corrections, $\delta_{\rm TPE}$ are the nuclear structure corrections dominated by the two-photon exchange, and $\delta_{\rm FS}(r^2_d)$ is the finite size correction proportional to the deuteron charge radius $r_d$. The bottle-neck in the precise determination of $r_d$ are the nuclear structure corrections. In this work, we overview the process of the uncertainty quantification of $\delta_{\rm TPE}$ in $\mu$D using nucleon-nucleon (NN) potentials at various orders (from LO to N$^4$LO) in chiral effective field theory (EFT).

\section{Analysis of uncertainties}
To quantify the total theoretical uncertainties of $\delta_{\rm TPE}$, all relevant uncertainty sources must be identified and estimated \cite{Hernandez_2018,Ji_2018}. These various sources are:

\begin{itemize}
\item  $\sigma_{\rm stat}$ : uncertainties arising from the spread of the low-energy constants (LECs) $\tilde{\alpha}$ in the nuclear potential;
\item $\sigma_{\rm T^{\rm Max}_{\rm Lab}}$ : uncertainties from the maximum lab energy $T^{\rm Max}_{\rm Lab}$ used in the fits of the NN potential;
\item $\sigma_{\rm \Delta}$ : uncertainty due to the truncation of the chiral order;
\item $\sigma_{\rm \Lambda}$ : uncertainty from the variations of the the cut-off $\rm \Lambda$ in the NN potentials;
\item $\sigma_{\eta}$ : uncertainty due to the expansion (on a parameter known as $\eta$) which we use in relating $\delta_{\rm TPE}$ to the nuclear response functions;
\item $\sigma_{\rm J}$ : uncertainties from systematic approximations in the electromagnetic operators $J^{\mu}(x)$;
\item $\sigma_{\rm N}$ : uncertainties due to single nucleon physics;
\item $\sigma_{Z\alpha}$ : uncertainties arising from higher $(Z\alpha)$ corrections.
\end{itemize}
For an observable $A$, the statistical uncertainties $\sigma_{\rm stat}(A)$ induced by variations in the LECs $\tilde{\alpha}$ of the NN potential are calculated around their optimal values $\tilde{\alpha}_0$ by assuming that the LECs follow a multivariate Gaussian probability distribution. Under these conditions the leading approximation to $\sigma_{\rm stat}(A)$ will be given by
\begin{equation}
\sigma^2_{\rm stat}(A) = \langle A^2 \rangle -\langle A \rangle = \vec{J}_A {\rm Cov}(\tilde{\alpha}_0) \vec{J}^T_A ,
\end{equation}
where ${\rm Cov}(\tilde{\alpha}_0)$ represents the covariance matrix of the LECs at the optimum, and
$\vec{J}_A$ is the Jacobian vector of $A$ with respect to the LECs,
\begin{equation}
J_{A,i} = \left( \frac{\partial A}{ \partial \tilde{\alpha}_{i}}\right)_{\tilde{\alpha}=\tilde{\alpha}_0}.
\end{equation}
The systematic uncertainties $T^{\rm Max}_{\rm Lab}$ arise from the energy span in the NN scattering data used to fit the LECs. This uncertainty was estimated from the N$^k$LO$_{\rm sim}$ potentials ($k=0,1,2$) \cite{Carlsson_2016} by varying the maximum lab energies of the fit from 125 MeV to 290 MeV and their uncertainties $\sigma_{\rm T^{\rm Max}_{\rm Lab}}$ where found to dominate over the statistical uncertainties $\sigma_{\rm stat}$.

The chiral truncation uncertainties $\sigma_{\rm \Delta}$ originate from the calculation of an observable $A(p)$ at a finite order $\nu$, with associated momentum scale $p$. This observable is assumed to obey the same expansion as the underlying NN-force given by
\begin{equation}
A(p) = A_0\sum_{\mu=0}^\nu c_{\mu}(p)Q^{\mu},\label{eq:chiral_expansion}
\end{equation}
\begin{figure}[!htb]
\minipage{0.32\textwidth}
  \includegraphics[width=\linewidth]{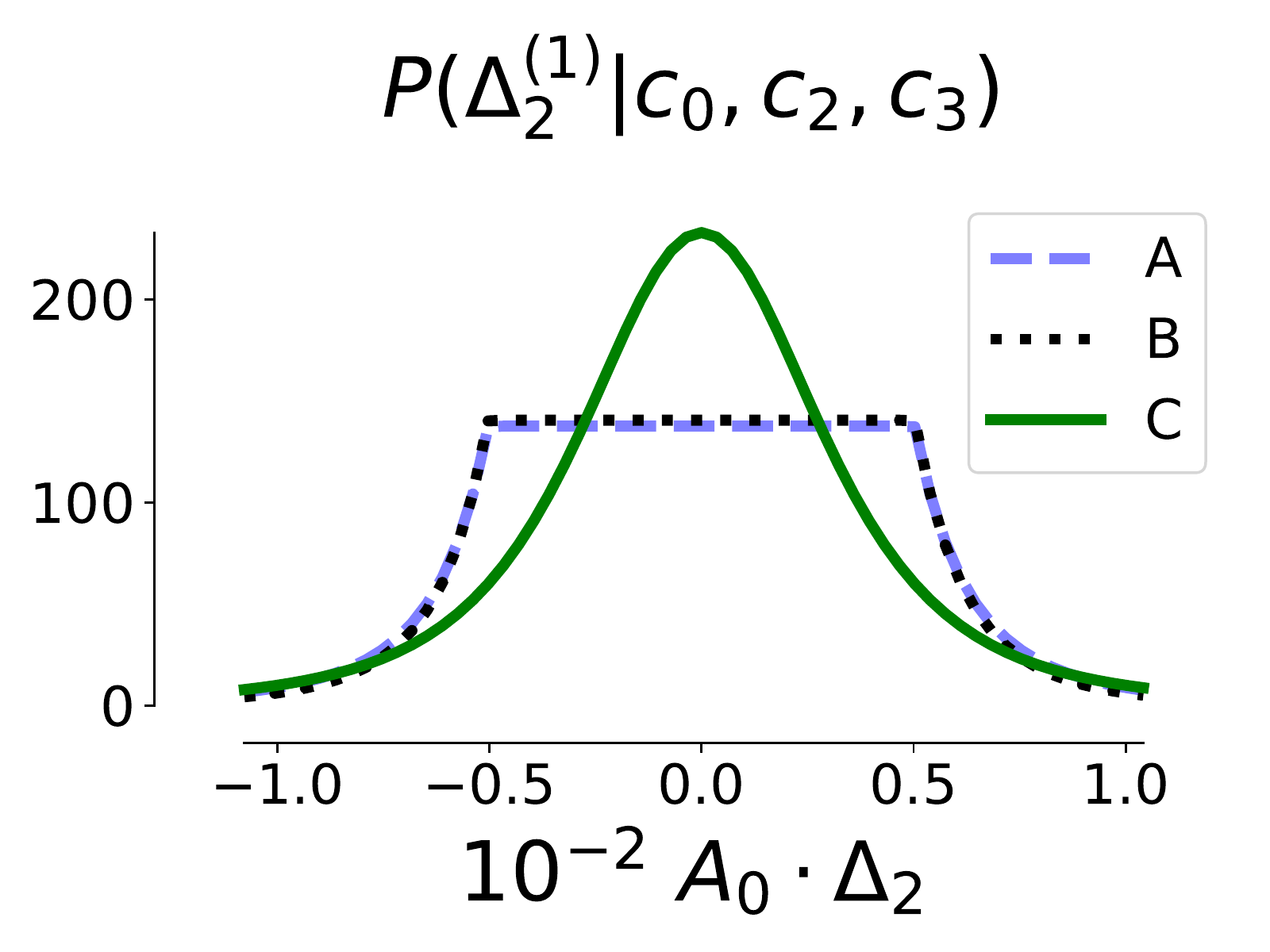}
\endminipage\hfill
\minipage{0.32\textwidth}
  \includegraphics[width=\linewidth]{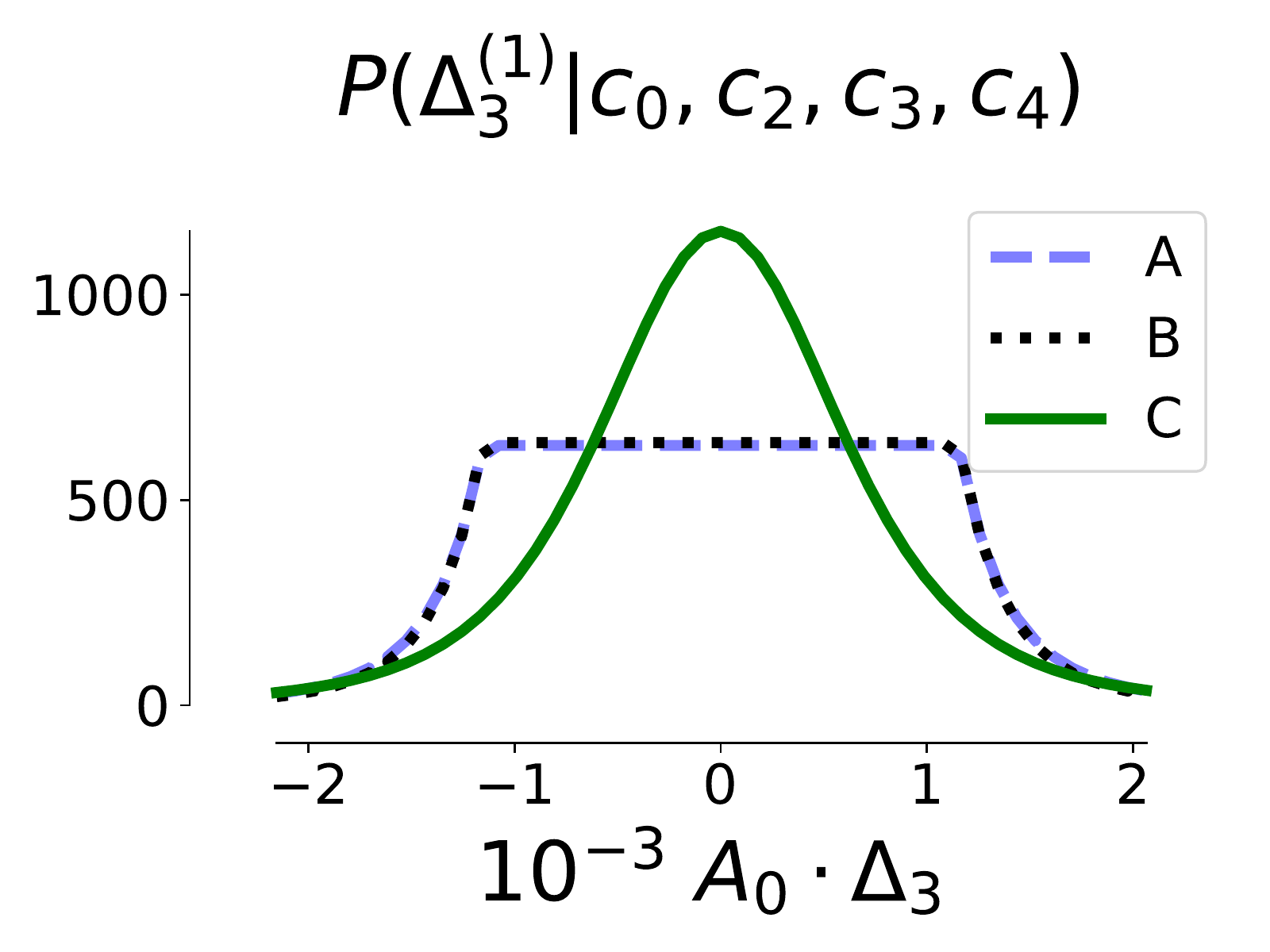}
\endminipage\hfill
\minipage{0.32\textwidth}%
  \includegraphics[width=\linewidth]{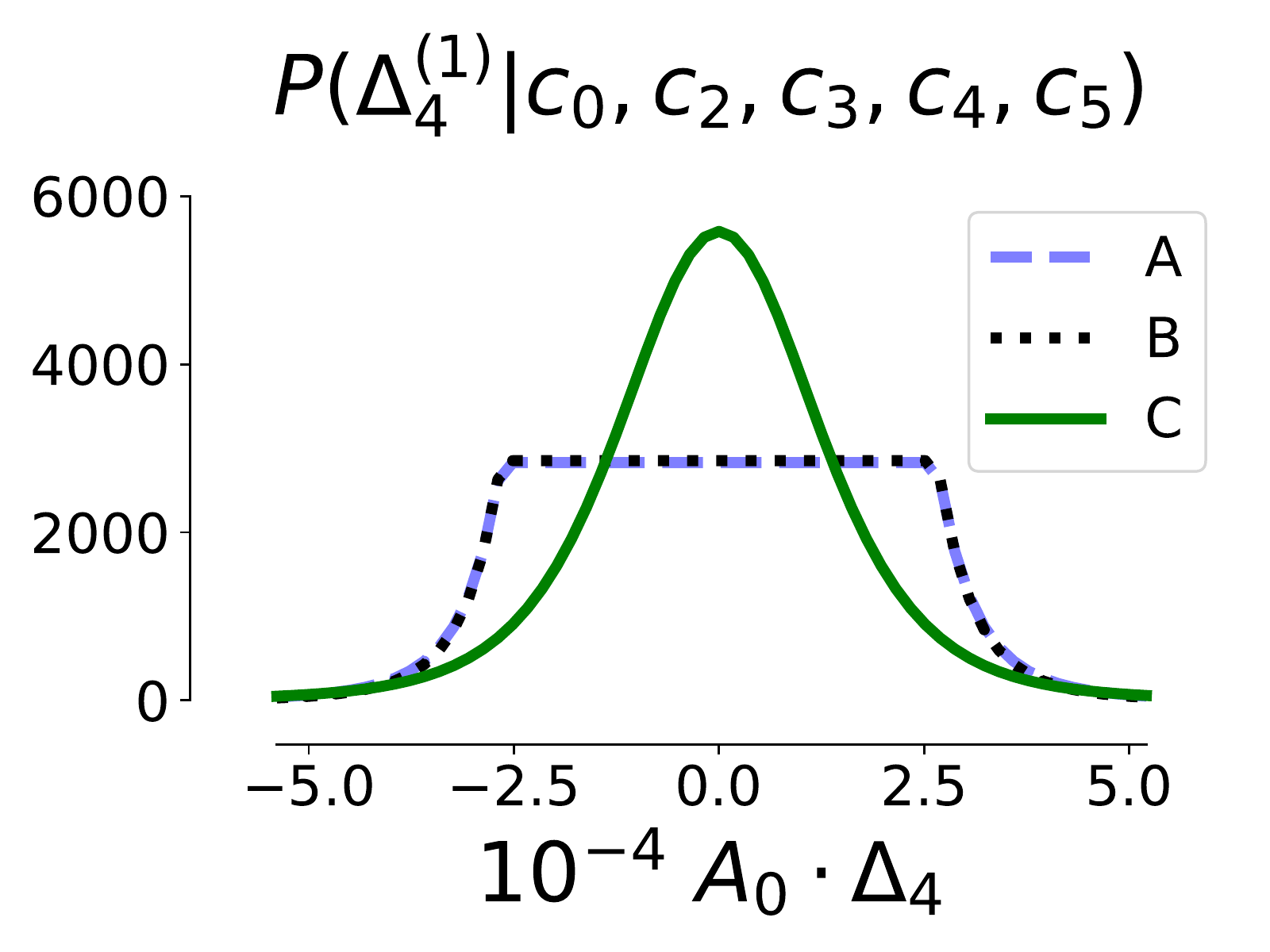}
\endminipage
\caption{Posterior distributions of the truncation uncertainties at different orders in the chiral EFT expansion ($A_{0}{\rm \Delta}^{(1)}_{2}$, $A_{0}{\rm \Delta}^{(1)}_{3}$, $A_{0}{\rm \Delta}^{(1)}_{4}$) in meV units for $\delta_{\rm TPE}$ in the leading-omitted term approximation using the potential in Ref.~\cite{Epelbaum_2015} with $(R_0, \Lambda)$=(0.8, 600) [fm, MeV]. The expansion parameter is $Q=0.23$, $\sigma$=1 for prior B, $\bar{c}_<$=0.1 and $\bar{c}_>$=10.}\label{fig:posterior distributions}
\end{figure}

where $A_0$ is the result at leading order, $Q$ is the expansion parameter, and $c_{\mu}(p)$ are observable and interaction specific coefficients assumed to be independent and of natural size. Assuming that the next higher-order term ${\rm \Delta}^{(1)}_{\nu} \equiv c_{\nu+1}Q^{\nu+1}$ dominates the truncation uncertainty in the calculation of $A(p)$, then the Bayesian posterior $P({\rm \Delta}^{(1)}_\nu)$ is given by \cite{Furnstahl_2015}
\begin{equation}
P({\rm \Delta}^{(1)}_{\nu}) = \frac{\int d\bar{c} \ P(c_{\nu+1}|\bar{c})P(c_0|\bar{c})P(c_2|\bar{c})...P(c_\nu|\bar{c})P(\bar{c})}{Q^{\nu+1}\int d\bar{c} \ P(c_0|\bar{c})P(c_2|\bar{c})...P(c_\nu|\bar{c})P(\bar{c})}, \label{eq:bayesian truncation}
\end{equation}
where $P(c_\mu|\bar{c})$ is the distribution of $c_\mu$ conditioned on the scale parameter $\bar{c}$ and $P(\bar{c})$ is the prior. In this contribution we update the results in Ref.~\cite{Hernandez_2018} by evaluating the 68$\%$ confidence intervals of the posteriors given in Eq.~(\ref{eq:bayesian truncation}) that represent the chiral truncation uncertainty $\sigma_{\rm \Delta}$. The posterior distributions $A_0 {\rm \Delta}^{(1)}_\nu$ from N$^2$LO to N$^4$LO for $\delta_{\rm TPE}$ using the chiral potentials from Ref.~\cite{Epelbaum_2015} are given in Fig.~\ref{fig:posterior distributions} for the priors A, B, C from Table I in  Ref.~\cite{Furnstahl_2015}.

Along with chiral truncation uncertainties, the chiral NN-potentials carry a parameter $\rm \Lambda$ that regulates the interactions. The systematic uncertainties $\sigma_{\rm \Lambda}$ arising from the regulators was probed using multiple cut-off values in the calculations of $\delta_{\rm TPE}$. These variations were found to be more significant than the uncertainties due to the chiral truncation.

The $\eta$-expansion arises from the calculation of $\delta_{\rm TPE}$ as a power series of the dimensionless operator $\eta \ll 1$. In the work of Ref.~\cite{Hernandez_2018,Ji_2018}, this expansion was carried out to sub-sub-leading order in $\eta$ and the truncation uncertainty $\sigma_{\eta}$ from higher order terms was determined to be $0.3\%$.

Uncertainties from approximations in the electromagentic operators $J^{\mu}(x)$, were estimated from the dipole response functions of Arenh{\"o}vel \cite{Arenhovel_private} that included MEC and relativistic corrections. Both of these effects were of the order $0.05\%$.

The uncertainties $\sigma_{\rm N}$ from single nucleon contributions to the TPE are an input in our analysis and taken from Ref.~\cite{Krauth_2016,Carlson_2014} and Ref.~\cite{Hill_2017}. Lastly, there was an estimated $1\%$ uncertainty from higher order $(Z\alpha)^6$ corrections, that include the three photon exchange.

\begin{table}
\caption{The uncertainty breakdown of the $\delta_{\rm TPE}$ at N$^4$LO}\label{table:uncertainty quantification}
\begin{center}
\begin{tabular}{l@{\quad\quad}l@{\quad\quad}l}
\hline
Source & $\%$ Uncertainty& Uncertainty in meV\\\hline
$\sigma_{\rm syst}$  & \begin{tabular}{c}$+0.5$ \\$-0.6$\end{tabular} & \begin{tabular}{c}$+0.008$ \\$-0.011$\end{tabular} \\ \hline
$\sigma_{\rm stat}$  &   0.06 & $\pm$0.001  \\ \hline
$\sigma_\eta$  &  0.3& $\pm$0.005 \\ \hline
$\sigma_{\rm N}$  & 0.6~/~1.2 &$\pm$0.0102 \cite{Krauth_2016}~/~$\pm$0.0198 \cite{Hill_2017} \\
$\sigma_{Z\alpha}$  &  1.0 &  $\pm$0.0172 \\ \hline
$\sigma_{\rm Total}$ & 1.3~/~1.6-1.7 & \begin{tabular}{c}$+0.022$~/~$+0.028$ \\$-0.023$~/~$-0.029$\end{tabular}  \\[2pt]
\hline
\end{tabular}
\end{center}
\end{table}

\section{Results and Conclusions}
The results of the analysis outlined in the previous section are summarized in Table \ref{table:uncertainty quantification}. The systematic nuclear physics uncertainty $\sigma_{\rm syst}$ is a combination of the $\sigma_{\rm \Delta}$, $\sigma_{\rm J}$ and $\sigma_{\rm T^{\rm Max}_{\rm Lab}}$ uncertainties, while $\sigma_{\rm Total}$ is a quadrature sum of all items in Table \ref{table:uncertainty quantification}. The calculation of $\sigma_{\rm \Delta}$ through the explicit calculation of the 68$\%$ confidence interval of the Bayesian posteriors instead of the prescription in Ref.~\cite{Epelbaum_2015} increases the lower bound slightly in $\sigma_{\rm Total}$ from -0.024 meV  in Ref.~\cite{Hernandez_2018} to -0.023 meV when using the $\sigma_{\rm N}$ values of Ref.~\cite{Krauth_2016} since the values of $\sigma_{\rm \Delta}$ at N$^4$LO for prior A are smaller when computed this way. The final value for the TPE correction was taken to be the average value of the calculations at N$^4$LO yielding $\delta_{\rm TPE}= -1.715$ meV with the final uncertainty $\sigma_{\rm Total}$. This value differs from the experimentally determined value from Ref.~\cite{Pohl_2016} of $\delta_{\rm TPE}=-1.7638(68)$ meV by less than 2 $\sigma$, which is not significant. From Table \ref{table:uncertainty quantification} we find that the uncertainties arising from the nuclear model dependence, $\sigma_{\rm syst}$ and $\sigma_{\rm stat}$, are small in comparison to the higher order $\sigma_{Z\alpha}$ or $\sigma_{\rm N}$ \cite{Hill_2017} uncertainties which dominate the total uncertainty. It is therefore unlikely that any differences between the experimental and theoretical determinations of $\delta_{\rm TPE}$ stem from models of the NN-forces.  

%
%

\end{document}